\documentclass[prb,aps,twocolumn,reprint,amsmath,floatfix,amssymb,showpacs,superscriptaddress]{revtex4-1}
\usepackage{graphicx}
\usepackage{bm}

\makeatletter

\usepackage{babel}
\usepackage{hyperref}
\hypersetup{colorlinks,allcolors=blue}

\newcommand{\aru}{$\alpha$-RuCl$_3$ }
\newcommand{\kggp}{$K\Gamma\Gamma'$ }

\begin{document}
\preprint{APS/123-QED}
\title{Spin excitation continuum to topological magnon crossover and thermal Hall conductivity in Kitaev magnets}

\author{Emily Z. Zhang}
\affiliation{Department of Physics, University of Toronto, Toronto, Ontario M5S 1A7, Canada}

\author{Reja H. Wilke}
\affiliation{Department of Physics, Arnold Sommerfeld Center for Theoretical Physics (ASC), Ludwig-Maximilians-Universität München, 80333 München, Germany}

\author{Yong Baek Kim}
\affiliation{Department of Physics, University of Toronto, Toronto, Ontario M5S 1A7, Canada}

\date{\today}

\begin{abstract}
There has been great interest in identifying a Kitaev quantum spin liquid state in frustrated magnets with bond-dependent interactions. In particular, the experimental report of a half-quantized thermal Hall conductivity in \aru in the presence of a magnetic field has generated excitement as it could be strong evidence for a field-induced chiral spin liquid. More recent experiments, however, provide a conflicting interpretation advocating for topological magnons in the field-polarized state as the origin of the non-quantized thermal Hall conductivity observed in their experiments. An inherent difficulty in distinguishing between the two scenarios is the phase transition between a putative two-dimensional spin liquid and the field-polarized state exists only at zero temperature, while the behaviour at finite temperature is mostly crossover phenomena. In this work, we provide insights into the finite temperature crossover behavior between the spin excitation continuum in a quantum spin liquid and topological magnons in the field-polarized state in three different theoretical models with large Kitaev interactions. These models allow for a field-induced phase transition from a spin liquid (or an intermediate field-induced spin liquid) to the field-polarized state in the quantum model. We obtain the dynamical spin structure factor as a function of magnetic field using molecular dynamics simulations and compute thermal Hall conductivity in the field-polarized regime. We demonstrate the gradual evolution of the dynamical spin structure factor exhibiting crossover behaviour near magnetic fields where zero-temperature phase transitions occur in the quantum model. We also examine nonlinear effects on topological magnons and the validity of thermal Hall conductivity computed using linear spin wave theory. We discuss the implications of our results to existing and future experiments.
\end{abstract}

\maketitle

\section{Introduction}
A tremendous effort has been made in identifying quantum spin liquid (QSL) states in real materials due to their ability to host fractionalized excitations and emergent gauge fields\cite{witczak-krempa_correlated_2014,savary_quantum_2016,zhou_quantum_2017,broholm_quantum_2020}. One way to achieve a type of QSL state, known as the Kitaev spin liquid (KSL), is through bond-dependent interactions $J_{\text{eff}}=1/2$ moments on a honeycomb lattice\cite{kitaev_anyons_2006, takagi_concept_2019}.
There has been substantial recent experimental developments and debate surrounding $\alpha$-RuCl$_3$\cite{plumb__2014,sandilands_scattering_2015,banerjee_proximate_2016,banerjee_neutron_2017,baek_evidence_2017,do_majorana_2017,wang_magnetic_2017,banerjee_excitations_2018,ye_quantization_2018,vinkler-aviv_approximately_2018,kasahara_majorana_2018,ye_phonon_2020,yokoi_half-integer_2021, czajka_oscillations_2021,zhou_intermediate_2022}, a candidate KSL material possessing a large ferromagnetic Kitaev interaction. At zero field, \aru is magnetically ordered due to the presence of non-Kitaev interactions\cite{rau_generic_2014,sears_magnetic_2015,johnson_monoclinic_2015,winter_2017_breakdown,winter_challenges_2018,winter_probing_2018}. The main controversy, however, involves the possibility of a field-induced quantum spin liquid state\cite{kasahara_majorana_2018,czajka_oscillations_2021,yokoi_half-integer_2021,czajka_planar_2022}. In the pure Kitaev model, a small magnetic field will induce a chiral spin liquid with Majorana fermion edge modes, leading to a half-quantized thermal Hall effect\cite{kitaev_anyons_2006}. Recent experiments reported a half-quantized thermal Hall conductivity when the magnetic field is applied in the in-plane direction\cite{yokoi_half-integer_2021}. Since then, however, there have been conflicting reports whose data appear to be more consistent with topological magnons arising in the field-polarized state\cite{czajka_oscillations_2021,czajka_planar_2022}.

From a theoretical point of view, a sufficiently high magnetic field leads to the field-polarized state which hosts topological magnons\cite{mcclarty_topological_2018,chern_sign_2021,zhang_topological_2021}. Thus, an important question remains about how one could distinguish between a putative intermediate-field spin liquid and the field-polarized state. A prominent difficulty in distinguishing between these two phases at finite temperature is that the phase transition between a possible two-dimensional spin liquid and the field-polarized state only exists at zero temperature\cite{kitaev_anyons_2006,yoshitake_majorana-magnon_2020}. Hence, the finite temperature behaviour near the transition would be a crossover phenomenon. For example, the spin continuum -- a hallmark of a quantum spin liquid\cite{knolle_dynamics_2014,knolle_dynamics_2015} -- may crossover to topological magnons as the magnetic field increases. The central questions are therefore how such a crossover would occur in the spin excitation spectrum, and what one should expect in the thermal Hall conductivity at finite temperature. 

In this work, we provide important insight into these questions by studying the crossover between the spin excitation continuum and topological magnons in frustrated magnets with bond-dependent interactions. In previous studies of the pure Kitaev models, it was shown that the dynamical structure factor containing a spin excitation continuum in the quantum model can be faithfully represented by the molecular dynamics (MD) result of the corresponding classical model\cite{samarakoon_comprehensive_2017,samarakoon_classical_2018}. This correspondence occurs because the momentum and energy dependencies of the continuum are mostly determined by the structure of the degenerate ground state manifold. Thus, we use MD to characterize the crossover behaviour in various theoretical models with a dominant Kitaev interaction. We then compute the thermal Hall conductivity of the field-polarized state using linear spin wave theory (LSWT), and investigate the limits of its validity and the causes of its breakdown.  

We investigate three different models with a magnetic field along the perpendicular direction to the honeycomb plane. All three models possess a zero temperature phase transition from a spin liquid (or a disordered state) to the field polarized states as a function of magnetic field. First, we consider the pure ferromagnetic Kitaev model, where there exists a direct transition from the chiral spin liquid to the field-polarized state at zero temperature. In this case, a quantum Monte Carlo result of the dynamical structure factor for the quantum model is available\cite{yoshitake_majorana-magnon_2020}. We benchmark our MD results against these quantum results, and demonstrate excellent agreement, validating our approach. Next, we consider the pure antiferromagnetic Kitaev model, where exact diagonalization (ED)\cite{hickey_emergence_2019} and density matrix renormalization group (DMRG) computations\cite{zhu_robust_2018} find an intermediate spin liquid phase between a chiral spin liquid at low field and the polarized state at high field. This intermediate spin liquid was proposed to be a gapless $U(1)$ spin liquid state\cite{hickey_emergence_2019}. Finally, we study a realistic \kggp model for \aru\cite{rau_generic_2014}, where there exists intermediate phases between the low-field zig-zag magnetic order and the field-polarized state at high field\cite{chern_magnetic_2020,gohlke_emergence_2020}. Note that extensive studies of realistic models for \aru did not find an intermediate quantum spin liquid when a magnetic field was applied along the in-plane direction\cite{sears_phase_2017,balz_finite_2019,balz_field-induced_2021}, which is at odds with experimental reports of a half-quantized thermal Hall conductivity\cite{yokoi_half-integer_2021}. On the other hand, various theoretical works find that a $c$-axis magnetic field (see Fig. \ref{fig:honeycomb}) leads to an intermediate quantum disordered state\cite{gohlke_emergence_2020}, which has recently been reported to exist in pulsed magnetic field experiments on \aru\cite{zhou_intermediate_2022}. In ED, DMRG, and Tensor Network studies of the quantum model, this intermediate field regime is identified as a possible quantum spin liquid\cite{gordon_theory_2019,gohlke_emergence_2020}. For this reason, we focus on the effect of a $c$-axis magnetic field. 

In general, we find distinctive finite temperature crossover behaviors in the dynamical spin structure factor near the magnetic field where a zero temperature transition from a spin liquid to the field-polarized state is seen in the quantum model. When a field-induced intermediate spin liquid (or a putative spin liquid) is present in the quantum model, we see redistribution of spectral weight in the dynamical spin structure factor. We then compute thermal Hall conductivity in the field polarized state using the linear spin wave theory and investigate its evolution near the phase boundary to a spin liquid regime. It is found that there exists significant non-linear effects in the magnon spectrum in the crossover region. Details of these crossover behaviors and thermal Hall conductivity are discussed in the main text.

The rest of the paper is organized as follows. In Sec. \ref{sec:methods}, we present the model and a description of the numerical methods used. Sections \ref{sec:kitaev} and \ref{sec:KGGp} presents the dynamical structure factor and thermal Hall conductivity results for the pure Kitaev and $K\Gamma\Gamma'$ models, respectively. Sec. \ref{sec:fielddep} presents the magnetic field dependence of the thermal Hall conductivity for all three models. Lastly, Sec. \ref{sec:discussion} discusses the key findings of this work and provides a future outlook.

\section{Model and Methods}\label{sec:methods}

\begin{figure}
    \centering
    \includegraphics[width=1.0\columnwidth]{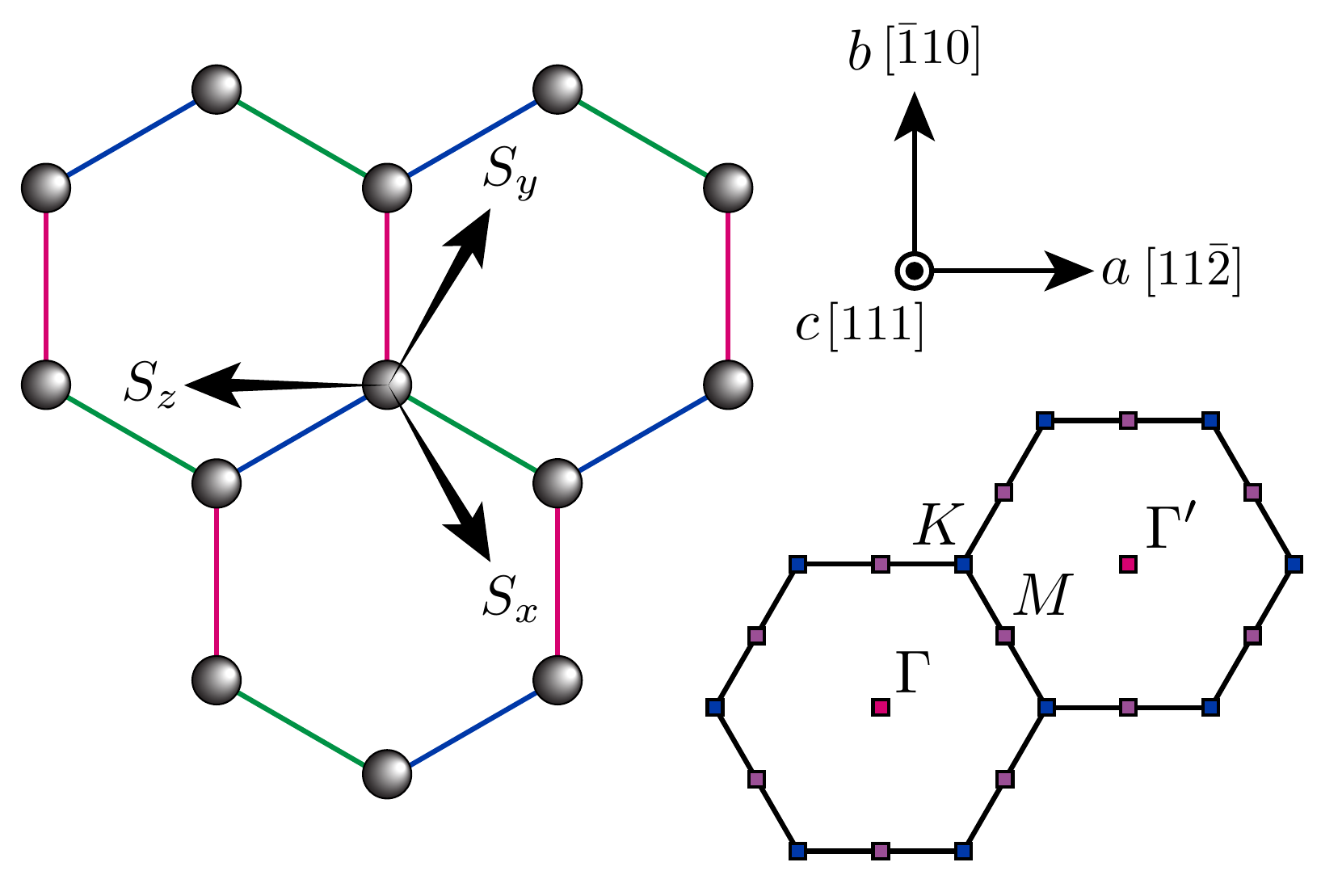}
    \caption{The $x$, $y$, and $z$ bonds of the Kitaev model are coloured in blue, green, and pink, respectively. The local spin axis ($S_x,S_y,S_z$) is shown coming out of the plane of the honeycomb, and the crystallographic axis ($a,b,c$) is indicated in the basis of the spin axis. The high symmetry points of the first Brillouin zone $\Gamma$, $M$, and $K$ are denoted with pink, purple, and blue squares, respectively. }
    \label{fig:honeycomb}
\end{figure}

\begin{figure*}[t!]
    \centering
    \includegraphics[width=1.0\textwidth]{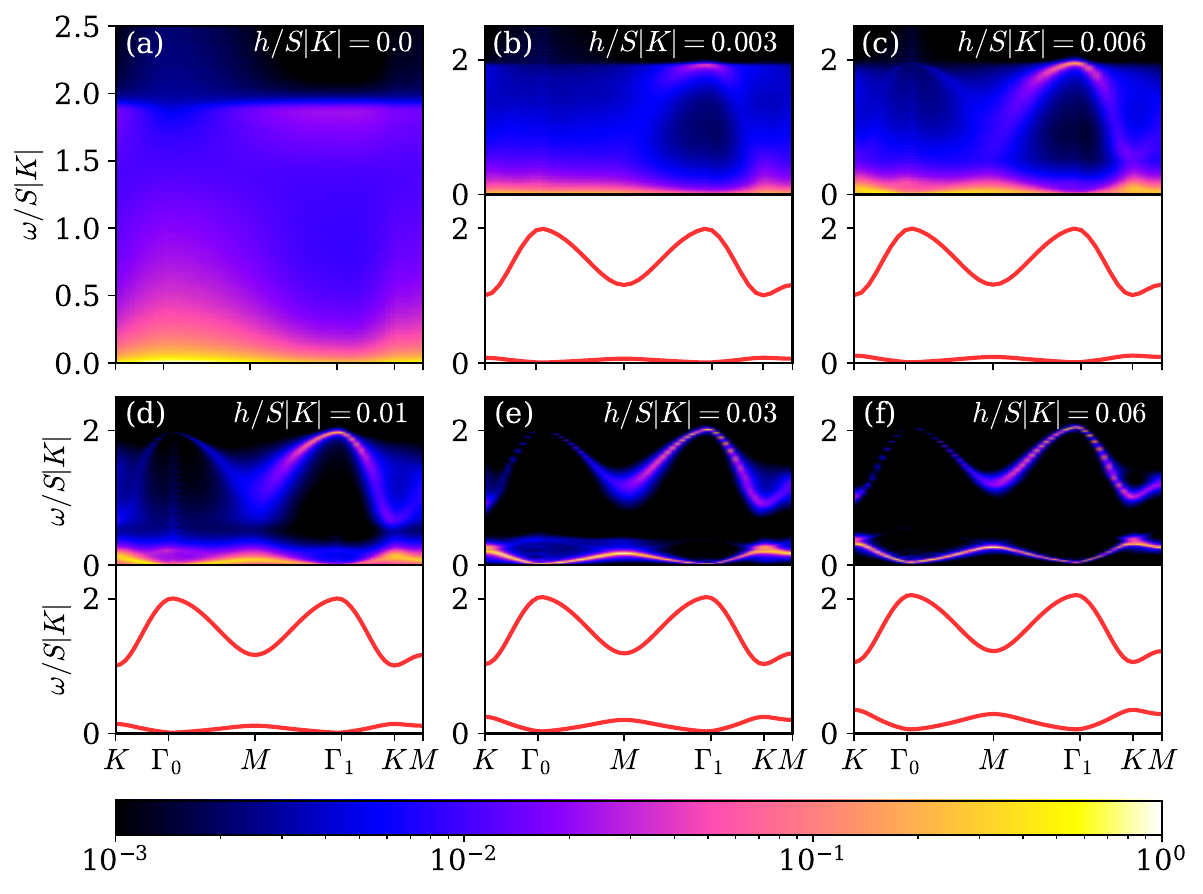}
    \caption{Field dependence of the neutron scattering dynamical structure factor $\mathcal{S}(\mathbf{q},\omega)$ obtained from molecular dynamics simulations for $K=-1, \Gamma=0, \Gamma'=0$ at $T/|K|=0.001$. The intensities are normalized with respect to the maximum value of each plot and the colourbar is presented on a logarithmic scale. The magnon bands of the polarized phase computed with LSWT are shown below the respective molecular dynamics results in (b)-(f). }
    \label{fig:FM_DSSF}
\end{figure*}

For general discussion, we consider the nearest-neighbour $K\Gamma\Gamma'$ Hamiltonian with a Zeeman coupling given by $H=\sum_{\langle ij\rangle\in\lambda}\mathbf{S}_i^TH_\lambda\mathbf{S}_j
-\mathbf{h}^T\sum_i\mathbf{S}_i$ where 
\begin{align}
    H_{x}=\left[\begin{matrix}K & \Gamma' & \Gamma'\\
                \Gamma' & 0 & \Gamma\\
                \Gamma' & \Gamma & 0
                \end{matrix}\right],
    H_{y}=\left[\begin{matrix}0 & \Gamma' & \Gamma\\
                \Gamma' & K & \Gamma'\\
                \Gamma & \Gamma' & 0
                \end{matrix}\right],
    H_{z}=\left[\begin{matrix}0 & \Gamma & \Gamma'\\
                \Gamma & 0 & \Gamma'\\
                \Gamma' & \Gamma' & K
    \end{matrix}\right],
\end{align}
and an out-of-plane magnetic field $\mathbf{h}\parallel[111]$ is applied (see Fig. \ref{fig:honeycomb}). We treat the spins as classical vectors $\mathbf{S}_i=(S^x_i, S^y_i, S^z_i)$ with a fixed magnitude $S$ and use finite temperature Monte Carlo techniques to take measurements of the thermally fluctuating spin configurations\cite{zhang_dynamical_2019}. For reference, we also provide the zero temperature phase diagrams compiled from various sources \cite{hickey_emergence_2019, chern_magnetic_2020} for the models used in this manuscript in Appendix \ref{zerotpd}.

To capture the spin excitation spectrum, we compute the dynamical spin structure factor (DSSF), a quantity directly comparable to inelastic neutron scattering experiments. The energy- and momentum-dependent spin correlations are defined as 
\begin{align}
    \mathcal{S}^{\mu\nu}(\mathbf{q}, \omega)=\frac{1}{2\pi N}\sum_{i,j}^{N}\int \mathrm{d}t\  e^{-i\mathbf{q}\cdot(\mathbf{r}_i-\mathbf{r}_j)+i\omega t}\langle S^\mu_i(t)S^\nu_j(0)\rangle
    \label{eq:dssf}
\end{align}
where $N$ is the number of lattice sites. We investigated the spectrum with unpolarized neutrons, described by 
\begin{align}
    \mathcal{S}(\mathbf{q}, \omega)=\frac{1}{2}\sum_{\mu,\nu}\left[\hat{z}_\mu\cdot\hat{z}_\nu-\frac{(\hat{z}_\mu\cdot \mathbf{q})(\hat{z}_\nu\cdot\mathbf{q})}{q^2}\right]\mathcal{S}^{\mu\nu}(\mathbf{q}, \omega).
\end{align}
Here, $\hat{z}_\mu$ are the basis vectors for the local Kitaev frame, shown in Fig. \ref{fig:honeycomb}. We compute Eq. \ref{eq:dssf} using the measurements from the finite temperature Monte Carlo calculations and perform molecular dynamics simulations on them\cite{lakshmanan_fascinating_2011,rackauckas_differentialequationsjl_2017,zhang_dynamical_2019,moessner_properties_1998}. Details of the numerical techniques can be found in Appendix \ref{MDapp}. 

\begin{figure*}[t!]
    \centering
    \includegraphics[width=1.0\textwidth]{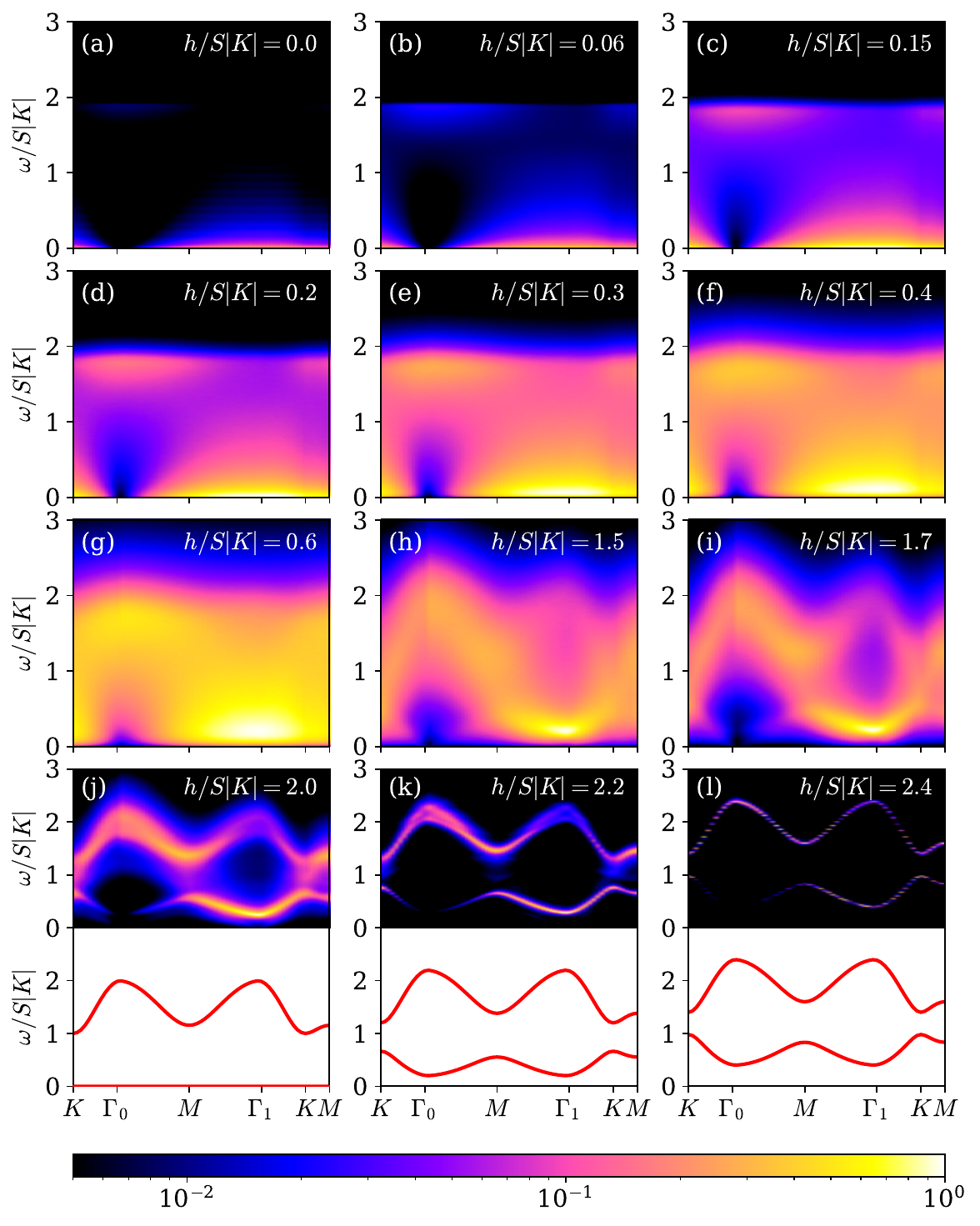}
    \caption{Field dependence of the neutron scattering dynamical structure factor $\mathcal{S}(\mathbf{q},\omega)$ computed with molecular dynamics for $K=1, \Gamma=0, \Gamma'=0$ at $T/|K|=0.001$. The intensities are normalized with respect to the maximum value of each plot and the colourbar is presented on a logarithmic scale. The magnon bands of the polarized phase computed with LSWT are shown below the respective molecular dynamics results in (j)-(l). The lower band in (j) occurs as a flat band at $\omega=0$.}
    \label{fig:AFM_DSSF}
\end{figure*}

We then calculate the thermal Hall conductivity using the framework of LSWT\cite{bogolyubov_theory_1947,matsumoto_theoretical_2011,matsumoto_thermal_2014,murakami_thermal_2017}, 
\begin{equation}
    \kappa_{xy}=-\frac{k_B^2T}{\hbar V}\sum_n\sum_{\mathbf{k}\in\text{FBZ}}\left\{c_2[g(\varepsilon_{n\mathbf{k}})]-\frac{\pi^2}{3}\right\}\boldsymbol{\Omega}_{n\mathbf{k}}
\end{equation}
where FBZ is the crystal first Brillouin zone, $c_2(x)=(1+x)[\ln(1+x)/x]^2-(\ln(x))^2-2\text{Li}_2(-x)$, Li is the dilogarithm, $g$ is the Bose-Einstein distribution, $\varepsilon_{n\mathbf{k}}$ is the dispersion of the magnon bands, and $\boldsymbol{\Omega}_{n\mathbf{k}}$ is the Berry curvature. $\kappa_{xy}/T$ was computed for the polarized state following the procedure described in \citep{chern_sign_2021,zhang_topological_2021}.

\section{Pure Kitaev Model}\label{sec:kitaev}
\subsection{Dynamical Spin Structure Factor}
First, we focus on the pure Kitaev limit with a ferromagnetic Kitaev coupling ($K<0, \Gamma=0, \Gamma'=0$). Figure \ref{fig:FM_DSSF} shows the evolution of the dynamical structure factor with increasing field strengths plotted along the FBZ path $K\to\Gamma_0\to  M\to\Gamma_1\to K\to M$ (see Fig. \ref{fig:honeycomb}). We can classify three distinct regimes to describe the crossover between the highly degenerate classical spin liquid state to the field-polarized state. Regime I occurs near $h=0$, where we observe a broad excitation profile with high intensities concentrated around $\Gamma_0$ at low energies, and a broad intensity at high energies near $\Gamma_1$. This regime is where the excitation continuum can be attributed to the fractionalization of the spins in the quantum model\cite{knolle_dynamics_2014,knolle_dynamics_2015}. In Regime II, $h$ is switched on and the spin continuum begins to show dispersive features, as seen in Figs. \ref{fig:FM_DSSF}(b)-(d). In this regime, a weak continuum coexists with weakly dispersing bands, and there are discrepancies when comparing with the magnon bands from LSWT. Lastly, the system is well polarized in Regime III, where the sharp bands from the DSSF agree well with LSWT, seen in Figs. \ref{fig:FM_DSSF}(e)-(f). Remarkably, the same qualitative behaviour can also be seen in quantum Monte Carlo results at finite temperature\cite{yoshitake_majorana-magnon_2020}. This agreement in the crossover behaviour for the FM Kitaev model thus validates our methodology when applying it to the next two models.

Next, we examine the antiferromagnetic Kitaev model  ($K>0, \Gamma=0, \Gamma'=0$) at varying field strengths in Fig. \ref{fig:AFM_DSSF}. We observe one crossover between the low-field chiral spin liquid and putative $U(1)$ spin liquid state, and another one as the system polarizes at high fields. We can identify four distinct regimes similar to the FM case, except with an additional intermediate regime corresponding to the $U(1)$ spin liquid. Figures \ref{fig:AFM_DSSF}(a)-(c) in Regime I shows a broad continuum with suppressed intensity at the $\Gamma_0$ and enhanced intensity at $\Gamma_1$ for the low energy mode, and a reversed intensity profile for the high energy mode. This regime is stable until approx. $h/S|K|=0.3$, where we begin to see a qualitative shift in the intensity distribution when crossing over to Regime II. Figures \ref{fig:AFM_DSSF}(d)-(g) demonstrate this progression in intensity shifts. Namely, we observe the broadening of the spectral weight at the high intensity points such that the continuum is almost uniform in intensity except for the small suppression at the $\Gamma_0$ point.  We note that the transition between Regimes I and II is not a sharp one, but rather a subtle redistribution of the spectral weight. Furthermore, Regime II corresponds to the region in the phase diagram obtained from exact diagonalization methods where there is a dramatic increase in the density of states when crossing from the KSL state to the gapless $U(1)$ spin liquid state\cite{hickey_emergence_2019}. This state was observed at similar field strengths in the quantum model, namely between approx. $h/S|K|=0.7$ and $1.2$, thus our MD results may apply in this region. Classically, this state is stable until approx. $h/S|K|=1.5$, where remnants of dispersive bands begin to appear in Figs. \ref{fig:AFM_DSSF}(h)-(j). This behaviour is characteristic of Regime III, where although bands are present in the DSSF, the system is not well-described by LSWT. Note that in Fig. \ref{fig:AFM_DSSF}(j), the lower magnon band is completely flat and gapless at $h/S|K|=2.0$. Finally, as the system fully polarizes at $h/S|K|=2.2$ in Regime IV shown in Figs. \ref{fig:AFM_DSSF}(k)-(l), we begin to observe good correspondence between the LSWT dispersion and MD calculations. 

\subsection{Thermal Hall conductivity}
\begin{figure}[h!]
    \centering
    \includegraphics[width=1.0\columnwidth]{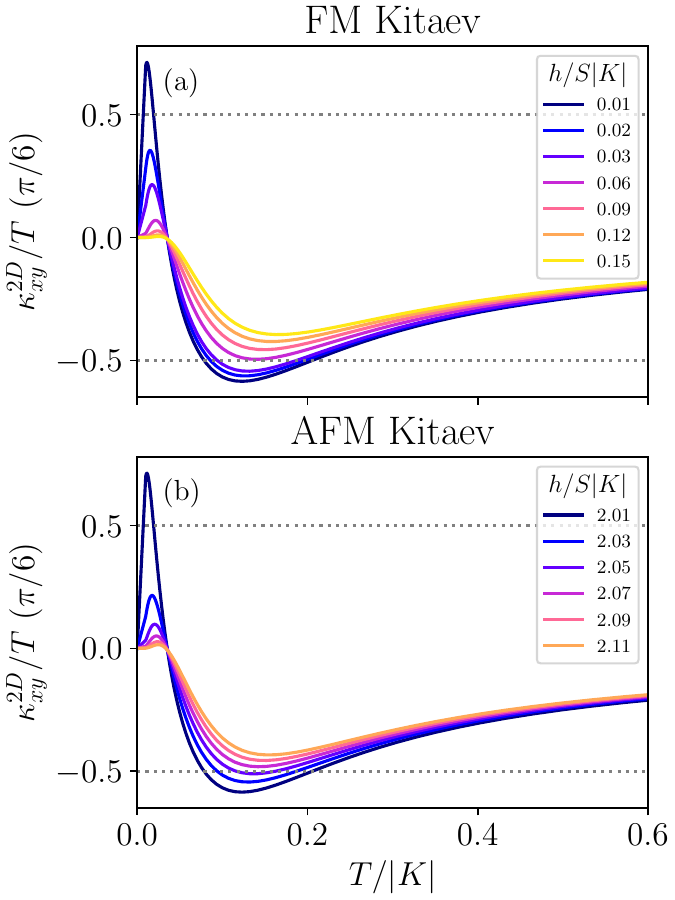}
    \caption{Two-dimensional thermal Hall conductivity $\kappa_{xy}^{2D}/T$ as a function of temperature due to magnons in the polarized state. $\kappa_{xy}^{2D}/T$ is in units of $\pi/6$, and we set $k_B=\hbar=1$ here. (a) was computed with interaction parameters $(K,\Gamma,\Gamma')=(-1,0,0)$ and (b) with $(K,\Gamma,\Gamma')=(1,0,0)$, both under a field $\mathbf{h}=h(1,1,1)/\sqrt{3}$. The half quantized values are indicated with the grey dashed line. }
    \label{fig:thc_kitaev}
\end{figure}
We present the thermal Hall conductivities due to magnons at different field strengths for the FM and AFM Kitaev model in Fig. \ref{fig:thc_kitaev}. The field strengths were chosen to be near the phase boundary of the polarized phase. Note that the FM case is consistent with \citet{mcclarty_topological_2018}. We first emphasize that for both cases, the magnitude of the thermal Hall conductivity can peak above the half-quantized value, especially close to the crossover regimes described above. In other words, we are able to achieve similar or equal magnitudes of $\kappa_{xy}^{2D}/T$ using only magnons in the polarized phase, and the actual magnitudes are heavily dependent on the parameter choice. We also note that the system polarizes at much higher fields for the AFM Kitaev model than FM Kitaev, which is consistent with the quantum phase diagrams\cite{hickey_emergence_2019}.

\begin{figure*}[t!]
    \centering
    \includegraphics[width=1.0\textwidth]{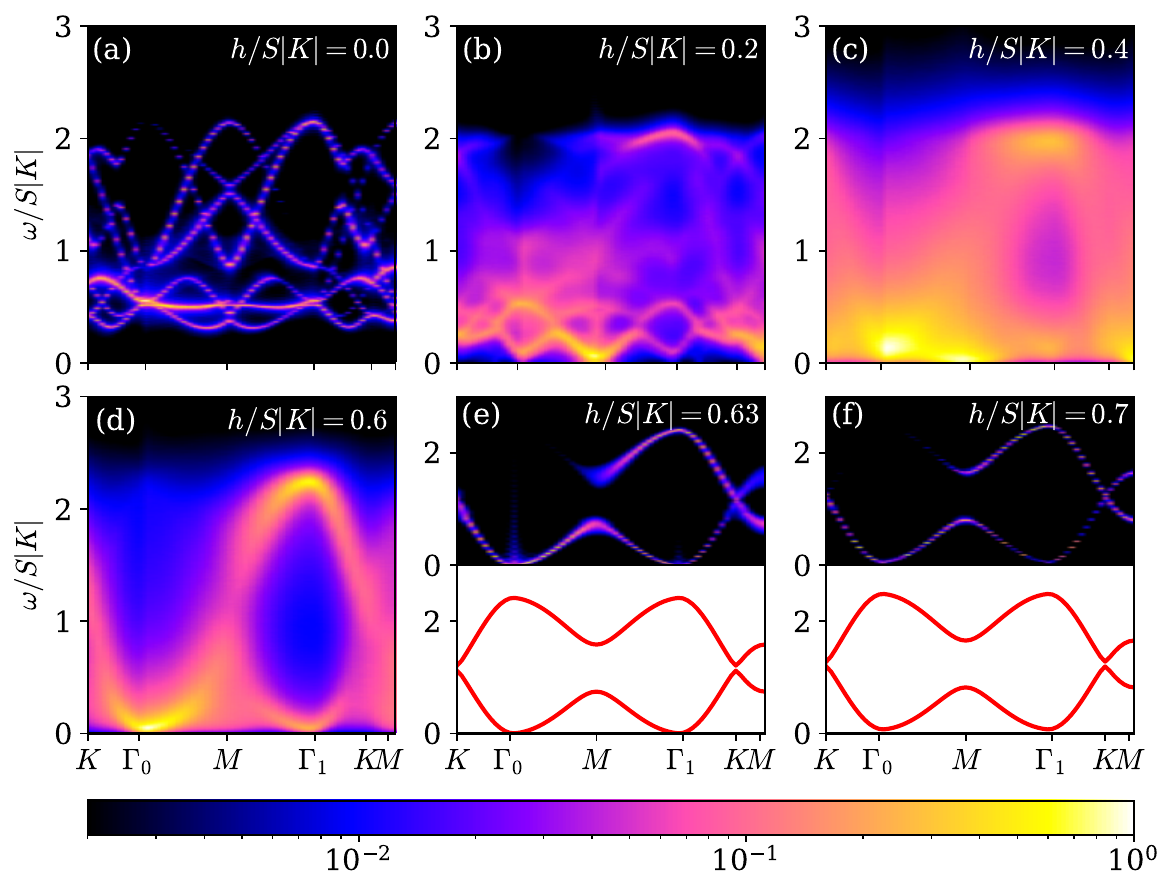}
    \caption{Field dependence of the neutron scattering dynamical structure factor $\mathcal{S}(\mathbf{q},\omega)$ computed with molecular dynamics for $K=-1, \Gamma=0.25, \Gamma'=-0.02$ at $T/|K|=0.001$. The intensities are normalized with respect to the maximum value of each plot and the colourbar is presented on a logarithmic scale. The magnon bands of the polarized phase computed with LSWT are shown under the respective molecular dynamics results in (e)-(f). }
    \label{fig:kggp_DSSF}
\end{figure*}

We further observe that there is a sign change in $\kappa_{xy}$ in both Figs. \ref{fig:thc_kitaev}(a) and (b), and the magnitude of $\kappa_{xy}$ is large at fields close to the critical field and at low temperatures. In the regions where the LSWT and MD results do not agree, $\kappa_{xy}$ computed within the framework of LSWT may not be reliable, especially at low temperatures. For example, in Fig. \ref{fig:thc_kitaev}(a), $\kappa_{xy}$ becomes larger at low temperatures for fields below $h/S|K|=0.06$, which coincides with the crossover Regime II for the FM Kitaev case described above. In this region, the magnitude of $\kappa_{xy}^{2D}$ becomes larger that 0.5, which may be renormalized with the inclusion of nonlinear effects. Similarly, for fields lower than $h/S|K|=2.2$, coinciding with crossover Regime III for the AFM Kitaev case in the MD calculations, $\kappa_{xy}$ becomes large below the temperature at which the sign change occurs. It is therefore in the crossover regions where we believe that higher order non-linear effects may need to be taken into consideration when computing the thermal Hall conductivity. 

\section{$K\Gamma\Gamma'$ model}\label{sec:KGGp}

Recent experiments using a pulsed magnetic field in the $c$-axis determined a quantum disordered phase at intermediate fields\cite{zhou_intermediate_2022}. To study this state, we examine the $K\Gamma\Gamma'$ model as a minimal model for \aru with an out-of-plane magnetic field. We choose an experimentally relevant parameterization $(K,\Gamma,\Gamma')=(-1,0.25,-0.02)$, where the classical ground state is magnetically ordered with large unit cells at intermediate fields between the zig-zag and polarized phases\cite{chern_magnetic_2020}. Since the classical large unit cell magnetic orders are very close in energy, these phases form a thermal ensemble at finite temperature. Figure \ref{fig:kggp_DSSF} shows the progression of the DSSF at increasing field strengths. At zero field in Fig. \ref{fig:kggp_DSSF}(a), we see a combination of sharp bands corresponding to the three configurations of the zig-zag order that arises from the $C_3$ symmetry of the honeycomb. As the field is switched on, we see the appearance of a large amount of bands in Fig. \ref{fig:kggp_DSSF}(b) corresponding to the frustration between several large unit cell magnetic orders. These bands eventually blur into continuum-like excitations seen in Figs. \ref{fig:kggp_DSSF}(c) and (d). In this regime, the large unit cell magnetic orders form a thermal ensemble at finite temperature and behave as the degenerate manifold for the intermediate state found in DMRG and tensor network studies\cite{gohlke_emergence_2020}. The blurring of the bands therefore corresponds to a quantum paramagnet; whether this state is indeed a spin liquid has yet to be determined in the study of the quantum model. Finally, the system crosses over to the polarized state as shown in Figs. \ref{fig:kggp_DSSF}(e) and (f). The behaviour of this crossover to the polarized state is similar to those observed in the FM and AFM Kitaev models, in that there are regions well-described and not well-described by LSWT. These results are consistent with a previous work that computed the dynamical structure factor using a stochastic Landau-Lifshitz approach that introduced finite temperature effects as thermal noise\cite{franke_thermal_2022}.
\begin{figure}
    \centering
    \includegraphics[width=1.0\columnwidth]{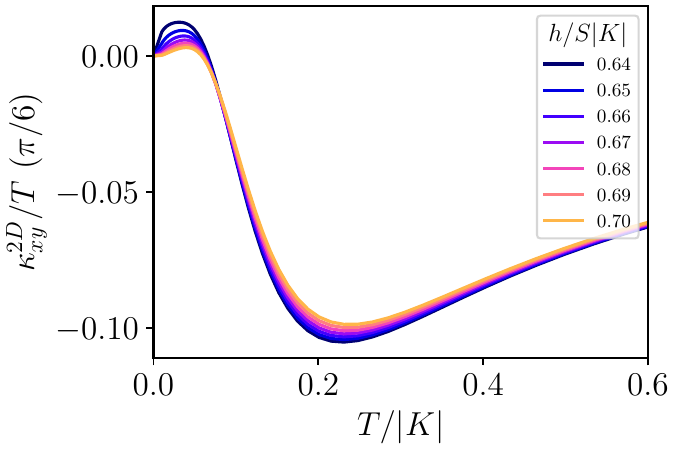}
    \caption{Two-dimensional thermal Hall conductivity $\kappa_{xy}^{2D}/T$ as a function of temperature due to magnons in the polarized state of the \kggp model. }
    \label{fig:thc_kggp}
\end{figure}
The thermal Hall conductivity due to magnons in the polarized state is presented for the \kggp model in Fig. \ref{fig:thc_kggp} for fields near the crossover. Although the peak of the thermal Hall conductivity is smaller than the half-quantized value, the qualitative behaviour is the same as the pure Kitaev models. Similar to the discussion for the pure Kitaev model, the results for $\kappa_{xy}$ are less reliable near the crossover region where nonlinear effects are strong, especially at low temperature. At higher fields where there is good agreement between LSWT and MD, $\kappa_{xy}$ computed with non-interacting magnons is sufficient in describing the magnon thermal Hall conductivity.

\section{Magnetic field dependence of the thermal Hall conductivity}\label{sec:fielddep}

\begin{figure}
    \centering
    \includegraphics[width=1.0\columnwidth]{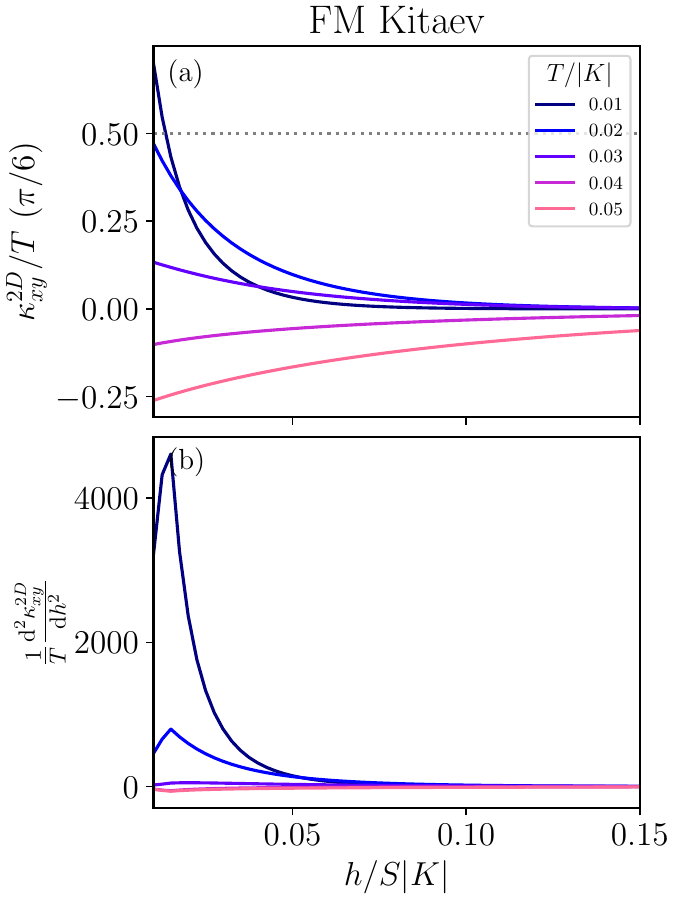}
    \caption{$\kappa_{xy}^{2D}$ vs $h/S|K|$ (a) and $\frac{1}{T}\frac{\text{d}^2\kappa_{xy}^{2D}}{\text{d}h^2}$ (b) for $K=-1$ for various temperatures. }
    \label{fig:FMfield}
\end{figure}

\begin{figure}
    \centering
    \includegraphics[width=1.0\columnwidth]{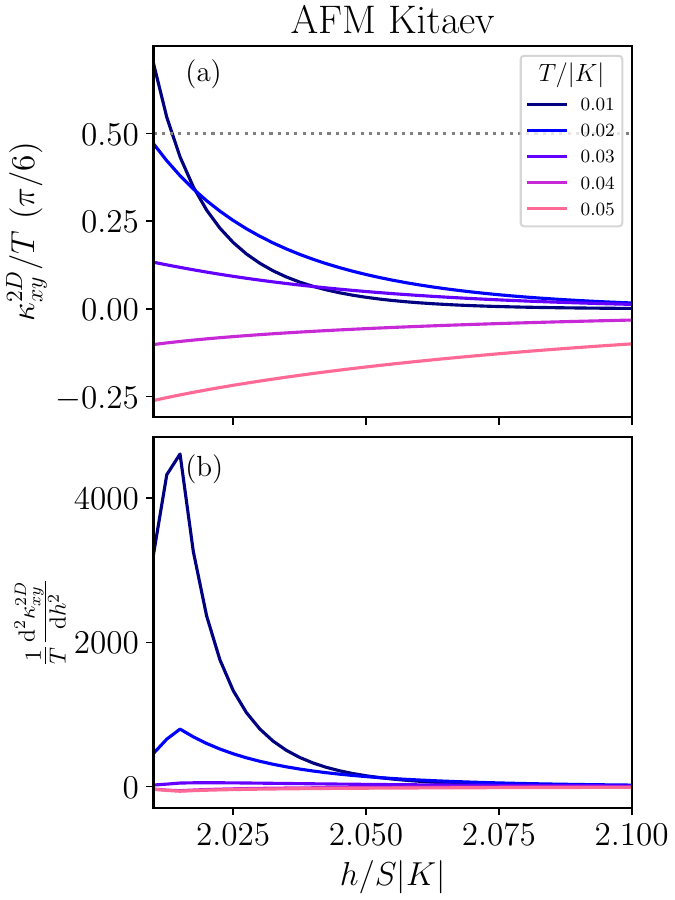}
    \caption{$\kappa_{xy}^{2D}$ vs $h/S|K|$ (a) and $\frac{1}{T}\frac{\text{d}^2\kappa_{xy}^{2D}}{\text{d}h^2}$ (b) for $K=1$ for various temperatures. }
    \label{fig:AFMfield}
\end{figure}

\begin{figure}
    \centering
    \includegraphics[width=1.0\columnwidth]{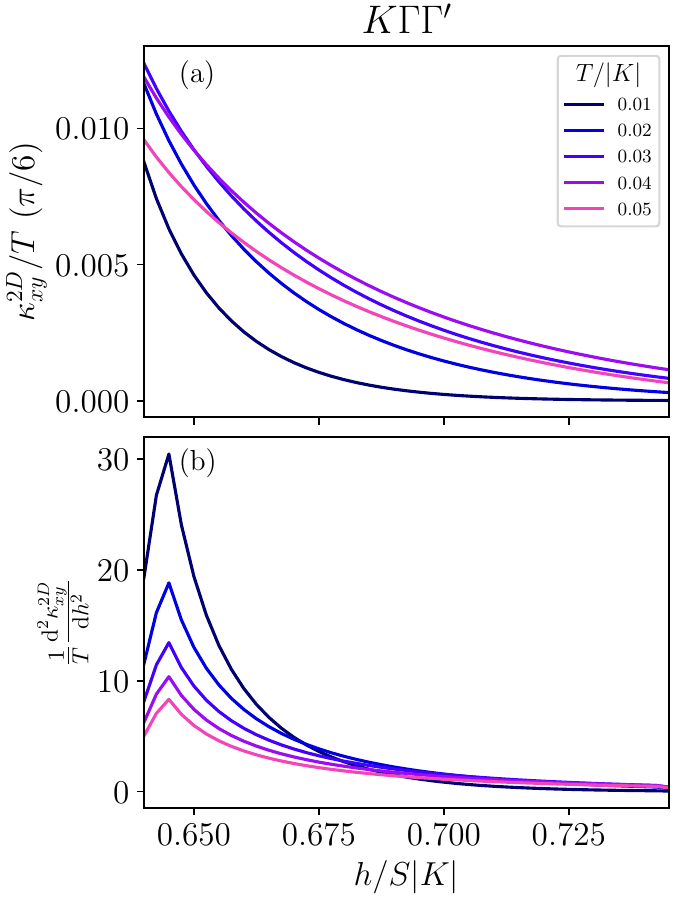}
    \caption{$\kappa_{xy}^{2D}$ vs $h/S|K|$ (a) and $\frac{1}{T}\frac{\text{d}^2\kappa_{xy}^{2D}}{\text{d}h^2}$ (b) for $K=-1, \Gamma=0.25, \Gamma'=-0.02$ for various temperatures. }
    \label{fig:KGGpfield}
\end{figure}

We present the thermal hall conductivity as a function of field along with their second derivatives in Figs. \ref{fig:FMfield}-\ref{fig:KGGpfield}. As made apparent in the second derivatives at low temperatures in all three models, we see a sharp peak at fields corresponding to a crossover between regimes. The peak positions in the second derivatives specifically occur at the crossover field strength between the regimes where there is a discrepancy between MD/LSWT and the agreement shown in earlier plots in the paper. For example, the peak in the FM Kitaev case at $h/S|K|=0.01$ coincides with Fig. \ref{fig:FM_DSSF}(d), where a continuum becomes apparent on top of the dispersive magnon bands. A similar comparison can be made between Fig. \ref{fig:AFMfield}(b) and \ref{fig:AFM_DSSF}(j) for the AFM Kitaev case, and between Fig. \ref{fig:KGGpfield}(b) and Fig. \ref{fig:kggp_DSSF}(e) for the \kggp case. Thus, in addition to the disagreement between MD and LSWT, these peaks in the second derivatives serve as an additional signature of the crossover between regimes where nonlinear effects become important, corroborating the above results.

\section{Discussion}\label{sec:discussion}

We first summarize our main results.
I) In the pure ferromagnetic Kitaev model at finite temperature, we investigated the evolution of the dynamical spin structure factor as a function of magnetic field and demonstrated the gradual crossover from the spin excitation continuum to topological magnons. II) In the pure antiferromagnetic Kitaev model, we found two crossovers in the dynamical spin structure factor at finite temperature, which is consistent with two transitions at zero temperature in the quantum model\cite{hickey_emergence_2019}. The spectral intensity distribution of the spin excitations are shown to be quite different between the low-field chiral spin liquid regime and the intermediate-field putative $U(1)$ spin liquid regime. Although many candidate Kitaev materials possess a ferromagnetic Kitaev coupling, recent studies have unveiled the potential realization of an antiferromagnetic Kitaev interaction in $f$-electron honeycomb materials\cite{jang_antiferromagnetic_2019,ishikawa_sm_2022},  polar spin-orbit Mott insulators\cite{sugita_antiferromagnetic_2020}, and $d^7$ compounds such as Na$_3$Co$_2$SbO$_6$ and Na$_2$Co$_2$TeO$_6$\cite{winter_magnetic_2022}. Whether these materials are proximate to the KSL or putative $U(1)$ spin liquid at remains to be explored. III) In the \kggp model, we again showed roughly two different crossover regimes marking the transition from the zigzag order to an intermediate field regime of high frustration (a putative spin liquid state may exist here in the quantum model), and then the transition to the field polarized state. IV) In the pure ferromagnetic and antiferromagnetic Kitaev models, we showed that the thermal Hall conductivity in the field polarized state can be larger than the half-quantized value. Interestingly, the peak value of the thermal Hall conductivity is close to the half-quantized value near the crossover regime. In the \kggp model, the peak value of the thermal Hall conductivity is smaller than the half-quantized value, although it depends on the precise values of the spin exchange interactions. 

We note that the results for the thermal Hall conductivity due to magnons in the FM Kitaev case (Fig. \ref{fig:thc_kitaev}(a)) were presented in a previous work\cite{mcclarty_topological_2018}. \citet{mcclarty_topological_2018} also studied the high-field topological magnon dynamical structure factor at zero temperature using LSWT, nonlinear spin wave theory using a $1/S^2$ expansion, and DMRG. In addition to the consistency of our MD results with theirs at high fields, we emphasize that our work is able to describe the entire crossover regime at finite temperature, especially at lower fields where magnon-based descriptions become unstable. 

Comparing the MD result of the dynamical spin structure factor and linear spin theory, we showed that there is significant nonlinear effects on the lower magnon band of the topological magnon spectrum in the crossover regime, whereas LSWT works well in higher magnetic fields. In addition, we see signatures of this crossover in the second derivative of the field-dependent thermal Hall conductivity. $\kappa_{xy}^{2D}$ is computed from LSWT, and it is currently not known how nonlinear effects can be taken into account. 
Hence, it is conceivable that the very low temperature behaviour of the thermal Hall conductivity in the crossover regime (or close to a phase transition) would deviate from the prediction of LSWT. Indeed, in a recent experiment with in-plane magnetic field\cite{czajka_planar_2022}, where thermal Hall conductivity was fit using topological magnon contributions, the very low temperature part of the data deviates from the LSWT prediction when the system is close to the critical magnetic field for the transition to the zigzag ordered state. 

A recent experiment on $\alpha$-RuCl$_3$ with out-of-plane magnetic field found a novel intermediate-field phase before the system enters the field-polarized state\cite{zhou_intermediate_2022}. The crossover behavior from the spin excitation continuum in an intermediate-field frustrated regime (or a putative spin liquid regime) to topological magnon regime in the \kggp model with out-of-plane field may be directly relevant to this experiment. In this case, one may be able to see such crossover in THz optical spectroscopy while scattering experiments at high fields may be out of reach. 

An important question remains as to what the behaviour of the thermal Hall conductivity would be in the crossover regime between a putative intermediate-field quantum spin liquid and the high-field polarized state at finite temperature.
The difficulty of describing this crossover regime may be one of the reasons why there have been conflicting experimental results on $\alpha$-RuCl$_3$ and their interpretations. 
Hence, it will be very useful to develop a general theoretical framework for an unbiased computation of thermal Hall conductivity irrespective of the nature of underlying phases or excitations. Such a framework may give us an important clue as to how one should interpret thermal Hall conductivity data in the crossover region of the finite temperature phase diagram if an intermediate-field spin liquid state does exist.

\begin{acknowledgments}
We thank F\'elix Desrochers for helpful discussions. We acknowledge support from the Natural Sciences and Engineering Research Council of Canada (NSERC). E.Z.Z. was further supported by the NSERC Canada Graduate Scholarships-Doctoral (CGS-D). R.H.W. was further supported by the German Academic Exchange Service (DAAD). Y.B.K. was further supported by Simons Fellowship from the Simons Foundation, and Guggenheim Fellowship from the John Simon Guggenheim Memorial Foundation. All the computations were performed on the Cedar supercomputer cluster, hosted by WestGrid in partnership with the Digital Research Alliance of Canada. 
\end{acknowledgments}
\bibliography{refs}

\appendix

\section{Zero-temperature Phase Diagrams of the Pure Kitaev and \kggp Models}\label{zerotpd}
\begin{figure}
    \centering
    \includegraphics[width=1.0\columnwidth]{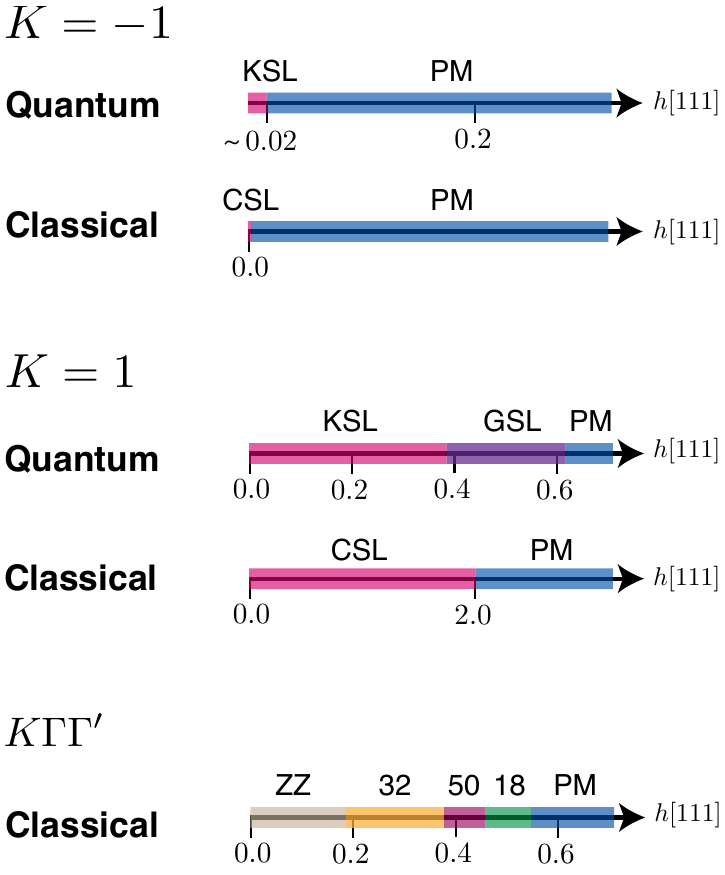}
    \caption{Zero-temperature phase diagrams as a function of field for the pure Kitaev and \kggp models. The phase labels correspond to the following: KSL=Kitaev spin liquid, PM=Polarized paramagnet, CSL=Classical spin liquid, GSL=Gapless $U(1)$ spin liquid, ZZ=zig-zag ordered phase, and 32, 50, and 18 correspond to 32-site, 50-site, and 18-site ordered phases respectively.}
    \label{fig:zeroT}
\end{figure}

The quantum and classical phase diagrams for $K=1$ and $K=-1$ are shown in Fig. \ref{fig:zeroT}. The quantum phase diagrams were obtained using exact diagonalization in \citet{hickey_emergence_2019}. We note that the transition between the Kitaev spin liquid phase and the gapless $U(1)$ spin liquid for $K=1$ is not observed classically at zero temperature. The phase diagram for $K=-1, \Gamma=0.25, \Gamma'=-0.02$ was obtained in  \citet{chern_magnetic_2020}
\section{Details of the Molecular Dynamics Calculation}\label{MDapp}

Firstly, we use finite temperature Monte Carlo (MC) techniques, specificially parallel tempering, to obtain the spin configurations needed to compute the spin correlations. We treat the spins classically, i.e. we treat the spins as vectors $\mathbf{S}=\left(S_{x}, S_{y}, S_{z}\right)$, and we fix the magnitude to be $S$. We study system sizes of up to $L=36\times 36\times 2$. We first perform at least $5 \times 10^{6}\ \mathrm{MC}$ thermalization sweeps. Then, we perform another $1 \times 10^{7} \mathrm{MC}$ measurement sweeps, with measurements recorded every 2000 sweeps. The spin configurations are then used as initial configurations (IC) for molecular dynamics (MD)\cite{zhang_dynamical_2019, moessner_properties_1998}, where each measurement is time-evolved deterministically according to the semi-classical Landau-Lifshitz-Gilbert equations of motion \cite{lakshmanan_fascinating_2011},
\begin{align}
\frac{d}{d t} \mathbf{S}_{i}=-\mathbf{S}_{i} \times \frac{\partial H}{\partial \mathbf{S}_{i}}.
\end{align}
The system evolves for up to $t|K|=700$, with step sizes of $\delta t|K|=0.05$ to obtain $S_{i}^{\mu}(t) S_{j}^{\nu}(0)$, in which the ICs are averaged over to obtain $\left\langle S_{i}^{\mu}(t) S_{j}^{\nu}(0)\right\rangle$. These results are then numerically Fourier transformed to obtain the momentum- and energy-dependent dynamical structure factors, $\mathcal{S}(\mathbf{q}, \omega)$. Our classical results are lastly re-scaled by a factor of $\beta\omega$, where $\beta=1/k_BT$, in order to reflect the classical-quantum correspondence $\mathcal{S}_{\mathrm{classical}}(\mathbf{q}, \omega)=\beta\omega\mathcal{S}_{\mathrm{quantum}}(\mathbf{q}, \omega)$ in the linear spin-wave theory framework \cite{zhang_dynamical_2019}. 

\end{document}